\documentclass[pre,twocolumn,showpacs]{revtex4}
\usepackage{amssymb}
\usepackage{graphicx}
\usepackage{amsmath}

\def\func#1{\mathop{\rm #1}\nolimits}
\def\Re{\func{Re}}

\begin{document}

\title{Theory of Diffusion Controlled Growth}
\author{R. C. Ball and E. Somfai}
\affiliation{Department of Physics, University of Warwick, Coventry CV4 7AL, U.K.}

\begin{abstract}
We expand upon a new theoretical framework for Diffusion Limited Aggregation
and associated Dielectric Breakdown Models in two dimensions [R.~C.~Ball and
E.~Somfai, Phys.~Rev.~Lett.~\textbf{89}, 135503 (2002)]. Key steps are
understanding how these models interrelate when the ultra-violet cut-off
strategy is changed, the analogy with turbulence and the use of logarithmic
field variables. Within the simplest, Gaussian, truncation of mode-mode
coupling, all properties can be calculated. The agreement with prior
knowledge from simulations is encouraging, and a new superuniversality of
the tip scaling exponent is discussed. We find angular resonances relatable
to the cone angle theory, and we are led to predict a new Screening
Transition in the DBM at large $\eta$.
\end{abstract}


\maketitle

\section{Introduction}

Diffusion controlled growth first attracted attention in the literature on
solidification, where the advance of a solidification front can be limited
by diffusion of either latent heat or compositional excess ahead of the
front. Under these conditions a planar front is linearly unstable with
respect to long wavelength corrugation, the Mullins-Sekerka instability
\cite{mullins63}, leading to a rich variety of problems in pattern formation.
Viscous fingering, arising when a viscous fluid is driven through a porous
medium by less viscous one, is recognised as being a problem in the same
class. The Diffusion Limited Aggregation model (DLA) of a rigid cluster
growing by the irreversible accretion of dilute diffusing particles,
introduced by Witten and Sander \cite{witten81} focussed attention on the
extreme limit of these problems, where all of the diffusion is ahead of the
growth and quasi-static, with the added simplification that the
Mullins-Sekerka instability applies on all lengthscales above the size of
the accreting particles.

Mathematically these problems share the same general form for the equations
governing their growth, with their local interfacial velocity controlled by
a conserved gradient flux, 
\begin{equation}
v_{n}\propto \left| \partial _{n}\phi \right| ^{\eta },\qquad \nabla
^{2}\phi =0,\text{\qquad }\phi _{\text{interface}}\approx 0  \label{growth}
\end{equation}
where [at least naively] $\eta =1$~\cite{ball86carg}. The generalisation to
a range of positive $\eta $ was introduced by Niemeyer, Pietronero and
Wiesmann \cite{niemeyer84} to model dielectric breakdown patterns, and in
our recent letter \cite{ball02prl} we introduced the idea that this can
support equivalences between models where the Mullins-Sekerka instability is
controlled locally in mathematically quite different ways.

The DLA model has attracted enormous attention because it contains no
limiting lengthscale (save the particle size) and so pattern formation must
continue non-trivially on all larger lengthscales, the Mullins-Sekerka
instability ruling out simple planar growth. Theoretical interest has been
fueled by the fractal and multifractal \cite{amitrano91,halsey86pra} scaling
properties of the clusters produced, with controversial claims \cite
{plischke84,coniglio89,mandelbrot02} (and counter-claims \cite
{meakin85,somfai99,ball02pre}) of anomalous scaling, and by the long-standing
absence of an overall theoretical framework to understand the problem. A
simple mean field theory \cite{ball84nw} does not capture the fractal
aspects, which are better understood through various relations between
exponents \cite{ball84w,turkevich85,ball86pa} and \cite{halsey87}. The Cone
Angle Theory \cite{ball86pa} gives a plausible argument for the fractal
dimension of DLA, whilst the Screened Growth Model \cite{halsey92} and
Makarov's Theorem \cite{makarov85} give insight into how the multifractal
spectrum of the growth is generated.

The presence of a cut-off lengthscale $a$ below which the physics dictates
smooth growth is a crucial ingredient of diffusion controlled growth; it is
known that otherwise infinitely sharp cusps develop in the interface within
finite time \cite{shraiman84}. In DLA this cutoff is fixed and set by the
size of accreting particles, but in solidification and also in viscous
fingering it is set by surface free energy leading to a local offset in the
value of $\phi _{\text{interface}}$: this is dependent on the local
interfacial curvature and leads to a velocity-dependent selection of
approximately constant $va^{2}$ for growing tips \cite{langer80}. We will
gather these different possibilities together with the generalised cut-off
law, that advancing tips have radius 
\begin{equation}
a\propto \left| \partial _{n}\phi \right| ^{-m}\text{.}  \label{tip}
\end{equation}
In terms of $m$, simple DLA corresponds to $m=0$ and solidification and
viscous fingering correspond to $m=1/2$; in the theory below in two
dimensions we will map onto the case where $a$ is such that each growing tip
has fixed integrated flux, corresponding to $m=1$.

This paper explains and expands the new theory announced in our recent letter 
\cite{ball02prl}. In Sec.~\ref{sec:mapping} we establish mappings between
models with different $\eta $ and $m$. This opens up the opportunity to
discuss the full class of models via the section at $m=1$, which we will
show in Sec.~\ref{sec:continuum-theory} is particularly amenable to
continuum theoretical description in two dimensions. Section \ref{sec:noise}
discusses the other key input to Sec.~\ref{sec:continuum-theory}, that we
regard diffusion controlled growth as a turbulence problem with
self-organising fluctuations. In Sec.~\ref{sec:renormalisation} we discuss
how to cast the theory of Sec.~\ref{sec:continuum-theory} in renormalised
form, with divergent factors factored out consistently. Closure
approximations are required to obtain explicit theoretical predictions, and
in Sec.~\ref{sec:gaussian-closure} we show how the simplest Gaussian closure
leads to a complete theory of the fractal and multifractal scaling. This
turns out to be quantitatively quite accurate for the zone of active growth.
In Sec.~\ref{sec:scaling} we show how one outstanding exponent, the tip
scaling of the harmonic measure, can be pinned down through the use of the
Electrostatic Scaling Law leading to a very surprising prediction with which
numerical data seems compatible. Section~\ref{sec:breakdown} gives a more
detailed discussion of what happens to the DBM exponents at large $\eta$. In
Sec.~\ref{sec:penetration-depth} we show examples of how the theory can be
deployed to tackle deeper quantities such as the relative penetration depth
in DLA, which has been the subject of several numerical studies and some
controversy. Angular resonances appear leading us to draw parallels in Sec.~%
\ref{sec:cone-angle-theory} with the earlier Cone Angle Theory.

\section{Scaling Properties and Mappings between models at different $%
\protect\eta $ and $m$}

\label{sec:mapping}

It is central to fractal (and multifractal) behaviour in DLA that the
measure given by the diffusion flux [density] $j\equiv\partial _{n}\phi $ onto
the interface has singularities \cite{halsey86pra}, such that the
integrated flux onto the growth within distance $r$ of a singular point is
given by 
\begin{equation}
\mu (r)\sim (r/R)^{\alpha },  \label{scaling}
\end{equation}
where $R$ is the overall linear size of the growth. Multifractal scaling of
the flux density leads to a whole spectrum of $\alpha $ values, with the
number of regions of lengthscale $r$ having $\mu (r)\sim (r/R)^{\alpha }$
varying as $(r/R)^{f(\alpha )}$, but in the following we focus particularly
on advancing tips and their associated exponent value $\alpha _{\text{tip}}.$

\begin{figure}[tbp]
\resizebox{0.9\columnwidth}{!}{\includegraphics*{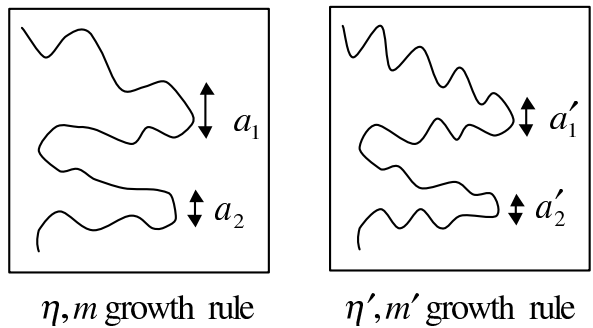}}
\caption{For equivalence between growth models with different values of $%
\protect\eta $ and $m$, we consider a pair of realisations matched down to
but not including tip radii. The equivalence then requires that the two
models agree about the relative velocities of competing tips.}
\label{fig:tipsfigure}
\end{figure}

Applying this phenomenology to the scaling around growing tips, we can
establish an equivalence between models at different $\eta $ and $m$ by
requiring that \textit{the relative advance rates of different growing tips
are matched}. Consider two growths, growing governed by different parameters 
$(\eta ,m)$ and $(\eta ^{\prime },m^{\prime })$ respectively, which at a
given moment have the same overall geometry down locally to the level of
[the coarser of] their cut-off lengthscales. For any given growing tip
(labelled $k$), the tip radius $a_{k}$ and flux density $j_{k}$ in the
unprimed growth will be related to those in the primed growth by $%
j_{k}a_{k}^{d-1}/a_{k}^{\alpha }=$ $j_{k}^{\prime }a_{k}^{\prime
d-1}/a_{k}^{\prime \alpha }$, where $\alpha $ is the local scaling exponent
(as per Eq.~(\ref{scaling})) of the harmonic measure between lengthscales $%
a_{k}$ and $a_{k}^{\prime }$. We take this exponent to have value $\alpha
=\alpha _{\text{tip}}$ on the grounds that this is locally a tip of the
growth. Now let us focus on two different growing tips labelled $1,2$ whose
radii and flux densities are interrelated in the unprimed growth according
to Eq.~(\ref{tip}) by $a_{1}j_{1}^{m}=a_{2}j_{2}^{m}$, and similarly in the
primed growth by $a_{1}^{\prime }j_{1}^{\prime m^{\prime }}=a_{2}^{\prime
}j_{2}^{\prime m^{\prime }}$. If we now insist that the advance velocities
are in the same ratio (between tips $1$ and $2$) in both models, this
requires $(j_{1}/j_{2})^{\eta }=(j_{1}^{\prime }/j_{2}^{\prime })^{\eta
^{\prime }}$, which forces the parameter relation 
\begin{equation}
\frac{1+m(1+\alpha -d)}{\eta }=\frac{1+m^{\prime }(1+\alpha -d)}{\eta
^{\prime }}.  \label{equivalence}
\end{equation}
For the two models to be equivalent in the relative velocities of all tips
requires their parameters be related as above, where $\alpha =\alpha _{\text{%
tip}}$ is the singularity exponent associated with growing tips.

\begin{figure}[tbp]
\resizebox{0.9\columnwidth}{!}{\includegraphics*{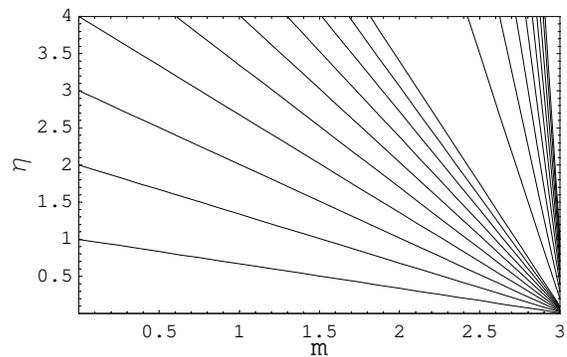}}
\caption{According to the mapping (\ref{equivalence}) models lying along any
given line shown are fundamentally related and should have equivalent
scaling properties. They should therefore be classifiable in terms of the
value $\protect\eta _{m}$ corresponding to any chosen reference value of $m,$
such as $\protect\eta _{0}$ or $\protect\eta _{1}.$ For simplicity the
graphic has been plotted for $d=2$ and taking $\protect\alpha =2/3$
completely independent of $\protect\eta $ (see later): if $\protect\alpha $
does vary from line to line, then the lines will not be confocal.}
\label{fig:mapping}
\end{figure}

Although we have not strictly proved the equivalence of the models related
above, we have shown that any such relationship must follow Eq.~(\ref
{equivalence}) and we will assume in the rest of this paper that this
equivalence holds. All such models are then classifiable in terms of a
convenient reference such as $\eta _{0}$, the equivalent $\eta $ when $m=0$,
corresponding to the original Dielectric Breakdown Model. For example
dendritic solidification with $\eta =1$ and $m=1/2$ corresponds to $\eta
_{0}=\frac{2}{3+\alpha -d}$: it is thus not equivalent to DLA, but to
another member of the DBM class.

Another puzzle resolved by our classification is a recent study showing
conflicting scaling between DLA and different limits of a ``Laplacian growth''
model \cite{barra01}. The latter model grows bumps of width proportional to
flux density, so in the present terminology it corresponds to $m=-1$. The
bumps are also grown with protrusion proportional to flux density. When
the coverage of the growing surface (per time step of growth) is low, then
as the bumps are also distributed proportional to flux density, this limit
corresponds to $\eta =3$. By contrast high coverage (with significant
suppression of overlapping bumps) corresponds to $\eta =1$. Using $\alpha
=0.7$ (see below) these map through Eq.~(\ref{equivalence}) into $\eta
_{0}=2.31$ and $\eta _{0}=0.77$ respectively, so the way their scaling was
observed \cite{barra01} to bracket that of DLA is quite expected.

\section{The Role of Noise}

\label{sec:noise}

DLA and DBM have been widely regarded as models in statistical physics, in
that the local advance rate in Eq.~(\ref{growth}) has been implemented as
the probability per unit time for the growth locally to make some unit of
advance, entailing an inherent shot noise. Here we argue that diffusion
controlled growth is a problem of turbulence type, with noise
self-organising from minimal input. This was first suggested by Sander et
al. \cite{sander85} but was only pursued in the case with surface tension
cut-off, $m=1/2$ in the present terminology, where it has been recognised
more recently as chaotic viscous fingering \cite{kessler01}.

\begin{figure}[tbp]
\resizebox{0.9\columnwidth}{!}{\includegraphics*{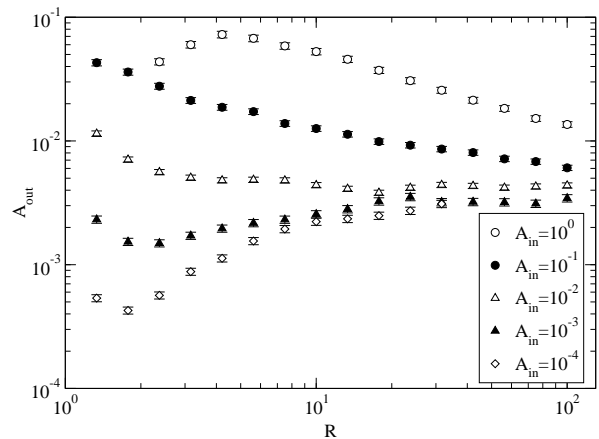}}
\caption{The size fluctuations (``output noise'') $A_{\text{out}}=(\protect%
\delta N/N)^{2}$, measured at fixed radius $R$ for DLA clusters grown with
off-lattice noise reduction \protect\cite{ball02pre} at various (``input'')
noise levels $A_{\text{in}}$. For low $A_{\text{in}}$, $A_{\text{out}}$ self
organizes from below in the manner of a turbulent system. }
\label{fig:noise}
\end{figure}

The renormalisation of noise with lengthscale has hitherto been discussed
(at least for DLA) in the context of noise reduction \cite
{barker90,cafiero93,ball02pre}, focussing on the idea that as one goes up in
lengthscale an equivalent coarse-scale model must have lower level of noise
than crude shot noise. The limiting or ``fixed-point'' level of noise in DLA
is small (at least according to references \cite{barker90} and \cite
{ball02pre}) but certainly not zero, so it is natural to ask whether it can
be approached from below as well as from above. The data in Fig.~\ref
{fig:noise} show clearly that for DLA grown with very low noise by the
methods of reference \cite{ball02pre}, the relative fluctuations do indeed
approach their limiting value from below as well as from above, and the same
result was implicit in the earlier Renormalisation Group results of \cite
{barker90}.

The above discussion leads us to conjecture that for the full range of
diffusion controlled growth under a continuum description of the interface,
disorder in the initial conditions alone should suffice to feed instability,
leading to the same limiting levels of structural fluctuation on larger
lengthscales as in the discrete models. The agreement we obtain below from
simulations of the continuum model without temporal noise provides direct
evidence for this idea, which might also be argued obvious on the grounds
that the Mullins-Sekerka instability \cite{mullins63} corresponds to a
preponderance of positive Lyaponov exponents in the dynamics.

\section{Continuum Theoretical Description and Self-Organisation}

\label{sec:continuum-theory}

The new ideas above, that we can balance changing the cut-off exponent $m$
by adjustment of $\eta $, and that noise can be left to self organise, are
the key to a new theoretical formulation of the problem, at least in two
dimensions of space to which we now specialise. In two dimensions the
Laplace equation in Eq.~(\ref{growth}) can be solved in terms of a conformal
transformation between the physical plane of $z=x+iy$ and the plane of
complex potential $\omega =\phi +i\theta ,$ in which we take the growing
interface to be mapped into the periodic interval $\theta =[0,2\pi ),\phi =0$
and the region outside of the growth mapped onto $\phi >0$. Then adapting
reference \cite{shraiman84}, Eq.~(\ref{growth}) leads to a closed equation
for the dynamics of just the interface $z(\theta )$ at $\phi =0$, 
\begin{equation}
\frac{\partial z(\theta )}{\partial t}=-i\frac{\partial z}{\partial \theta }%
\mathcal{P}\left[ \left| \frac{\partial \theta }{\partial z}\right| ^{1+\eta
}\right] .  \label{bseqn}
\end{equation}
The linear operator $\mathcal{P}$ is most simply described in terms of
Fourier transforms: $\mathcal{P}\left[ \sum_{k}e^{-ik\theta }f_{k}\right]
=\sum_{k}P(k)e^{-ik\theta }f_{k}=f_{0}+2\sum_{k=1}^{K}e^{-ik\theta }f_{k}$,
where we have introduced here an upper cut-off wavevector $K$. It is easily
shown that on scales of $\theta $ greater than $K^{-1}$ a smooth interface
is linearly unstable with respect to corrugation for $\eta >0$ (the
Mullins-Sekerka instability \cite{mullins63}), whereas for scales of $\theta 
$ less than $K^{-1}$ the equation drives smooth behaviour (corresponding
locally to the case $\eta =-1$). This cutoff on a scale of $\theta $, the
cumulative integral of flux, corresponds in terms of tip radii and flux
densities to $aj\thickapprox K^{-1}$, that is an $m=1$ cutoff law. Thus the
parameter $\eta $ in Eq.~(\ref{bseqn}) is more specifically $\eta
_{1}=\alpha \eta _{0}$, using Eq.~(\ref{equivalence}) with $d=2$.

We will present a numerical study of Eq.~(\ref{bseqn}) after a variable change
in Sec.~\ref{sec:scaling}, where disorder was supplied only through the
initial conditions. The results clearly confirm that the equation
self-organises critical scaling behaviour, without the supply of time
dependent noise. The surprising form of the scaling is interpreted below.

\section{Renormalisation and the DBM scaling law}

\label{sec:renormalisation}

We now turn to a theoretical analysis of Eq.~(\ref{bseqn}), and for
generality we will consider growth in a wedge of angle $2\pi c $
(with periodic angular boundary conditions) so that $c \rightarrow 0$
corresponds to growth along a channel whilst $c =1$ corresponds
``radial growth'', that is growth out from a point in the plane. The primary
theoretical requirement is that we must obtain results explicitly
independent of the cut-off as $K\rightarrow \infty $ : this is hard because
we will see that the mean advance rate of the interface diverges as a
power of $K$, and on fractal scaling grounds one would expect the same
divergent factor to appear in the rate of change of other simple variables.
One can of course take ratios of rates of change and look to order terms
such that divergences cancel. To make this work we have been forced to
introduce yet another change of variables, 
\begin{eqnarray}
-i\frac{\partial z}{\partial \theta } &=&c R\exp \left( i c
\theta -\lambda (\theta )\right)  \notag \\
&=&c R\exp \left( i c \theta -\sum_{k>0}
\lambda _{k}e^{-ik\theta }\right) ,  \label{logvars}
\end{eqnarray}
which corresponds to Fourier decomposing the logarithm of the flux density.
Here $R$ is the effective radius of the growth and the non-analytic factor
$e^{i c \theta }$ gives the mean winding of the conformal map through
the wedge angle $2\pi c $, leaving $\lambda (\theta )$ as a simple
Fourier series (except one-sided; see details in Appendix~\ref{app:logvar}).
The key to the success of the ``logarithmic variables''
$\lambda $ is that they decompose the flux density itself multiplicatively
and, as we shall see, quite naturally capture its multifractal behaviour. In
terms of these, time rescaled through $dt=(c R)^{2y}d\widehat{t}$ and 
$y=(1+\eta )/2$, the equation of motion (\ref{bseqn}) becomes 
\begin{multline}
\frac{d\lambda _{k}}{d\widehat{t}}=-\sum_{j<k}(k-j)\lambda _{k-j}P(j)\left(
e^{y(\lambda +\overline{\lambda })}\right) _{j} \\
+2(k-c )\left(
e^{y(\lambda +\overline{\lambda })}\right) _{k}  \label{logeom}
\end{multline}
where subscripts on bracketed expressions imply the taking of a Fourier
component, by analogy with $\lambda _{k}$. The advance rate of the mean
interface is correspondingly given by 
\begin{equation}
\frac{dR}{d\widehat{t}}=c R\left( e^{y(\lambda +\overline{\lambda }%
)}\right) _{0}.  \label{mean_velocity}
\end{equation}
Details of the above analysis are given in Appendix~\ref{app:logvar}.

At this point we can evaluate the multifractal spectrum in terms of these
logarithmic variables.  The multifractal spectrum of the harmonic measure
follows from computing the general moment \cite{halsey86pra} $Z(q,\tau )=\sum
\,\left| \delta \theta \right| ^{q}\,\left| \delta z\right| ^{-\tau }$, in the
limit where all the intervals $\left| \delta z\right| $ and correspondingly
$\delta \theta $ approach zero; then the locus $(q-1)D(q)=\tau (q)$
separates the limiting behaviour $Z(q,\tau )\rightarrow \infty $ from
$Z(q,\tau )\rightarrow 0$. In our case it is convenient to fix $\delta
\theta $ (admitting wide variations in $\left| \delta z\right| $) and we
must focus on the restricted range $K^{-1}<\delta \theta <1$.  For $\delta
\theta \simeq 1$ we have trivially $Z\simeq (cR)^{-\tau },$ whilst for $\
\delta \theta \simeq K^{-1}$ the growth begins to look smooth so we can
approximate $Z(q,\tau )\simeq \sum \,\left| \delta \theta \right| ^{q-\tau
}\,\left| \frac{\partial z }{\partial \theta}\right| ^{-\tau }\simeq
K^{-q+\tau +1}\,\int d\theta \left| \frac{\partial z}{\partial \theta}%
\right| ^{-\tau }$. Averaging gives $\left\langle \left| \frac{\partial
z}{\partial \theta}\right| ^{-\tau }\right\rangle =\left\langle (c
R)^{-\tau }e^{(\lambda +\bar{\lambda})\tau /2}\right\rangle$, yielding
\begin{equation}
Z(q,\tau )\simeq K^{-q+\tau +1}
(cR)^{-\tau }\left\langle e^{(\lambda +\bar{\lambda})\tau /2}\right\rangle
\label{Z}
\end{equation}
The multifractal spectrum is then readily obtained from the separator
behaviour discussed above, provided we can evaluate the average in
Eq.~(\ref{Z}).  In Sec.~\ref{sec:gaussian-closure} we show how this can be
done quite explicitly in the Gaussian Closure Approximation.

We can now derive an elegant combination of Halsey's Electrostatic Scaling Law
\cite{halsey87} combined with the tip scaling law of Ball and Witten
\cite{ball84w,turkevich85,ball86pa}, that for DLA
$\tau(3)=D_{f}=1+\alpha _{\text{tip}}$.  This new derivation (unlike
the earlier results) is not restricted to the $m=0$ case.
The key idea is that we
match the advance rate of the forward tips of the growth (governed by $\alpha
_{\text{tip}}$) to that of the mean radius governed by a multifractal moment
through Eq.~(\ref{mean_velocity}). For consistency with the rest of this paper
it is convenient to present the argument for the $m=1$ representation, leading
to tip velocity $\frac{dR}{dt}= j^{\eta_1}\approx \left( K^{-1}/a\right)
^{\eta _{1}}$ where the tip radius $a$ is set by the condition $K^{-1}\approx
aj \approx (a/R)^{\alpha _{\text{tip}}}$ and hence $\frac{dR}{dt}\approx
R^{-\eta _{1}}K^{\eta _{1}(1/\alpha_{\text{tip}}-1)}$.
The overall advance rate of the growth (in terms of its effective radius) is
given from Eq.~(\ref{mean_velocity}) by $\frac{dR}{dt}\approx (cR)^{-\eta_{1}}
\langle e^{y(\lambda +\bar{\lambda})}\rangle$.  For the special value
$\tau/2=y$ this can be substituted into Eq.~(\ref{Z}), giving
$\frac{dR}{dt}\approx R^{-\eta_1} K^{q(\tau)-\tau -1}$, when $Z$ is kept at its
$K$-independent value $(cR)^{-\tau}$.  Comparing the two results leads to the
DBM scaling law
\begin{equation}
q(\tau=1+\eta_1) = 2 + \eta_1/\alpha_{\text{tip}} = 2 + \eta_0 \,,
\label{electrostatic-scaling-law-qtau}
\end{equation}
or its inverse form
\begin{equation}
\tau(q=2+\eta_0) = 1 + \eta_1 = 1 + \alpha _{\text{tip}}\eta _{0} \,.
\label{electrostatic-scaling-law-eqn}
\end{equation}
The same result can be found, much more tortuously, from growth at general
$m$.

Now we turn back to the equation of motion of the logarithmic variables.
Let us suppose some ignorance of the initial conditions and describe the
system in terms of a joint probability distribution over the $\lambda _{k}$,
and let us denote averages over this [unknown] distribution by $\left\langle
..\right\rangle $. We can in principle determine the distribution through
its moments, whose evolution we now compute. For simplicity in this paper we
assume translational invariance with respect to $\theta $, so that only
moments of zero total wavevector need be considered, of which the lowest
gives: 
\begin{equation}
\begin{split}
\frac{d}{d\widehat{t}}\left\langle \lambda _{k}\overline{\lambda }
_{k}\right\rangle =\; & \bigg( -\sum\limits_{j<k}(k-j)P(j)\left\langle
\lambda _{k-j}\overline{\lambda }_{k}e_{j}^{y(\lambda +\overline{\lambda }
)}\right\rangle \\
&+2(k-c )\left\langle \overline{\lambda }_{k}e_{k}^{y(\lambda +
\overline{\lambda })}\right\rangle \bigg) +\left(\text{c. conj.}\right) .
\end{split}
\label{unrenormalisedEoM}
\end{equation}

All of the higher moments lead to the same form of averages on the RHS,
$\left\langle \text{multinomial(}\lambda ,\overline{\lambda }\text{)}%
e^{y(\lambda +\overline{\lambda })}\right\rangle $, and all of these terms
are conveniently expressed in terms of cumulants \cite{kubo62} as detailed
in Appendix~\ref{app:cumulant-expansion}. The key helpful feature is that the
expressions we require
all naturally divide by one factor of $\left\langle e^{y(\lambda +\overline{%
\lambda })}\right\rangle =\frac{1}{c }\frac{d}{d\widehat{t}}%
\left\langle \ln R\right\rangle $, which is what we need in order to
remove divergences by eliminating increment of time $d\widehat{t}$ in favour
of $\frac{1}{c }d\left\langle \ln R\right\rangle .$ The latter
quantifies the update of large scale geometry, in terms of the advance in
radius relative to the circumference of the wedge (or width of channel). The
evolution of the second moments is then given in renormalised form by 
\begin{equation}
\begin{split}
c \frac{d}{d\left\langle \ln R\right\rangle }\left\langle \lambda _{k}%
\overline{\lambda }_{k}\right\rangle =&\; \bigg( 2(k-c )\left\langle 
\overline{\lambda }_{k}e_{k}^{y(\lambda +\overline{\lambda })}\right\rangle
_{c} \\
&-\sum\limits_{j<k}(k-j)P(j) \bigg[ \left\langle \lambda _{k-j}\overline{%
\lambda }_{k}e_{j}^{y(\lambda +\overline{\lambda })}\right\rangle _{c} 
\\
&+\left\langle \lambda _{k-j}e_{j-k}^{y(\lambda +\overline{\lambda }%
)}\right\rangle _{c}\left\langle \overline{\lambda }_{k}e_{k}^{y(\lambda +%
\overline{\lambda })}\right\rangle _{c} \bigg] \bigg) \\
&+\left( \text{c. conj.}\right) 
\end{split}
\label{renormalisedEoM}
\end{equation}
where $\left\langle {}\right\rangle _{c}$ denotes the Kubo cumulant \cite
{kubo62}.

The above result (\ref{renormalisedEoM}) is the key analytical step in this
paper, because it removes divergent factors form the equations of motion. \
It is not dependent on the closure approximation discussed below, and
should support other approaches also. Moreover Eq.~(\ref{renormalisedEoM}),
with the hierarchy of analogous equations for the evolution of higher
moments, offers a new entry point towards exact results in the class of DLA
and DBM models.

\section{Gaussian Closure Approximation}

\label{sec:gaussian-closure}

To obtain simple tractable results we need to introduce some closure
approximation(s) and we present here the simplest, neglecting all cumulants
higher than the second, equivalent to assuming a joint Gaussian distribution
(of zero mean) for $\lambda $. This is entirely characterised by its second
moments $S(k)=\left\langle \lambda _{k}\overline{\lambda }_{k}\right\rangle $
which by Eq.~(\ref{logeom}) we find evolve according to 
\begin{equation}
c \frac{dS(k)}{d\left\langle \ln R\right\rangle }=2y^{2}S(k)\left( 
\frac{\eta _{1}}{y^{2}}(k-k^{\ast })-kS(k)-2\sum_{j<k}jS(j)\right)
\label{S(k) evolution}
\end{equation}
where $k^{\ast }=c (1+1/\eta _{1})$, and again the details are in
Appendix~\ref{app:cumulant-expansion}.

Equation (\ref{S(k) evolution}) evolves to a unique steady state.  The key to
understanding this is to note that for $k<k^{\ast }$ the whole factor in
large braces is negative definite, so $S(k)=0$ is the unique attractor. Then
for $k=k_{\text{min}}$, the first integer value above $k^{\ast }$, $S(k_{%
\text{min}})=0$ is unstable and the zero of the last factor leads to the
global attractor having $k_{\text{min}}S(k_{\text{min}})=$ $\frac{\eta _{1}}{%
y^{2}}(k_{\text{min}}-k^{\ast })$. The attractor values for higher $k$ now
follow by induction: denote the factor in large braces by $B(k)$ and assume
that the attractor has $B(k)=0$ and $0\leq kS(k)\leq $ $\frac{\eta _{1}}{%
y^{2}}$, which are true for $k=k_{\text{min}}$. Then it follows that $B(k+1)=%
\frac{\eta _{1}}{y^{2}}-kS(k)-(k+1)S(k+1)$ and for $kS(k)<\frac{\eta _{1}}{%
y^{2}}$ the attractor must in turn have $B(k+1)=0$ and hence a value of $%
S(k+1)$ conforming to $0\leq (k+1)S(k+1)\leq $ $\frac{\eta _{1}}{y^{2}}$. In
the case $kS(k)=\frac{\eta _{1}}{y^{2}}$ the only and stable solution is $%
S(k+1)=0$ which leads to the same conclusions. Thus by induction the only
stable attractor of the system has $B(k)=0$ for all $k>k^{\ast }$ and the
corresponding steady state values are 
\begin{alignat}{2}
kS(k) &=0,\quad & k&<k_{\text{min}}  \nonumber \\
kS(k) &=\frac{\eta _{1}}{y^{2}}(k_{\text{min}}-k^{\ast })\,,\quad &k&=k_{%
\text{min}},k_{\text{min}}+2,\dots \nonumber \\
kS(k) &=\frac{\eta _{1}}{y^{2}}(1+k^{\ast }-k_{\text{min}})\,,\quad &
k&=k_{\text{min}}+1,k_{\text{min}}+3,\dots \label{variances} 
\end{alignat}
Note that when $k^{\ast }$ is integer, as particularly in the case $c
=0$ corresponding to growth in a channel as discussed in \cite{ball02prl}, $%
k_{\text{min}}=1+k^{\ast }$ and alternate values of $kS(k)$ are zero: in
the case of channel growth this absence of even $k$ is readily interpreted
in terms of the dominance of one major finger and one major fjord.

Within the Gaussian approximation and its predicted variances
(\ref{variances}) we can now compute all [static] properties of diffusion
controlled growth. From Eq.~(\ref{Z}) we obtain $Z(q,\tau)\simeq (cR)^{-\tau
}\exp \left( \tau ^{2}/4\sum_{k}^{K}S(k)\right) \simeq R^{-\tau }K^{\frac{%
\tau ^{2}\eta _{1}}{8y^{2}}}$, using the values from Eq.~(\ref{variances}),
and hence $Z(q,\tau )\simeq R^{-\tau }K^{-q+\tau +1+\frac{\tau ^{2}\eta
_{1}}{8y^{2}}}$. Thus the separator of limiting behaviour (now as $%
K\rightarrow \infty $) is given by 
\begin{equation}
q(\tau )=1+\tau +\tau ^{2}\frac{\eta _{1}}{2\left( 1+\eta _{1}\right) ^{2}}
\label{moments}
\end{equation}
It is also easy to see that any closure scheme based on keeping cumulants of 
$\lambda $ up to some finite order leads to a corresponding degree
polynomial truncation of $q(\tau )$.

\begin{figure}[tbp]
\resizebox{0.9\columnwidth}{!}{\includegraphics*{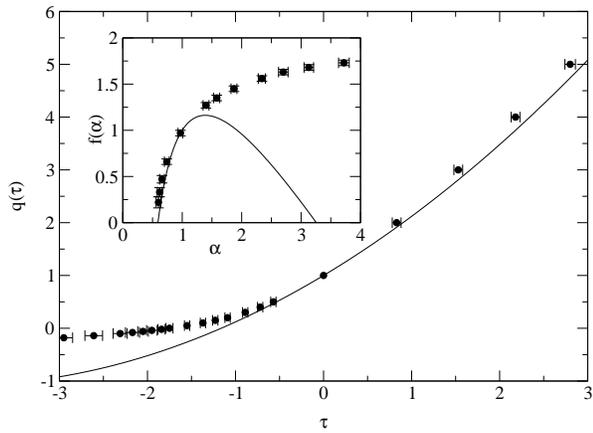}}
\caption{Multifractal spectra from the Gaussian theory
($\protect\alpha_\text{tip}=2/3$),
compared to measured values for DLA \protect\cite{ball90}. Agreement is
excellent for the active region $\protect\tau\geq 0$, $\protect\alpha\leq 1$%
, and there are no adjustable parameters. }
\label{fig:tauq}
\end{figure}

From the Legendre Transform of the inverse function $\tau (q)$ (as detailed in
Appendix~\ref{app:legendre}) we obtain the corresponding spectrum of
singularities, 
\begin{equation}
f(\alpha )=2-\frac{1}{\alpha }+\frac{1}{2}\left( \eta _{1}+\frac{1}{\eta _{1}%
}\right) \left( 2-\alpha -\frac{1}{\alpha }\right)  \label{falpha}
\end{equation}
which in Fig.~\ref{fig:tauq} is compared to measured data for DLA \cite
{ball90}, which later measurements \cite{jensen02} reinforce. For the region
of active growth $\alpha \leq 1$ ($q\geq 0$) the theory is quantitatively
accurate. At $\alpha =1$ it conforms to Makarov's theorem \cite{makarov85},
and in contrast to the Screened Growth Model \cite{halsey92} it does this
without adjustment. For $\alpha >1$ the spectrum is only qualitatively the
right shape, and for such screened regions our equations based on tip
scaling may not hold.

\section{Scaling predictions for DBM}

\label{sec:scaling}

To compare with the conventional DBM at $m=0$ and parameterised by $\eta
_{0} $ , we still need to compute theoretically the value of the tip scaling
exponent which enters through $\eta _{1}=\alpha _{\text{tip}}\eta _{0}$.

We can use the DBM scaling law (\ref{electrostatic-scaling-law-qtau}) with the
Gaussian Closure Approximation (\ref{moments}) for $q(\tau )$ to fix the value
of $\alpha _{\text{tip}}$, and the resulting
prediction is $\alpha _{\text{tip}}=2/3$ \emph{independent of} $\eta _{1}$.
For DLA in two dimensions this value is respectably close to (but outside)
measured values, $\alpha _{\text{tip}}=D-1=0.71\pm 0.01$ known from large
direct simulations of DLA \cite{ossadnik91,somfai000}, but its suggested
independence of $\eta $ over a range of DBM is quite shocking. Numerical
evidence, however, appears to lend support.

We have investigated numerically what value of $\alpha_{\text{tip}}$ is
seleced by the dynamics of Eq.~(\ref{bseqn}), with disorder
supplied only through the initial condition. Changing variables to
$\psi = \left(-i\frac{\partial z}{\partial \theta}(cR)^{-1}e^{-ic\theta}
\right)^{-(1+\eta _{1})/2} = e^{y\lambda}$, we obtain
\begin{equation}
\frac{\partial \psi }{\partial \widehat{t}}=
-i\frac{\partial \psi }{\partial \theta } \mathcal{P}[\psi \overline{\psi }]
+iy\psi \frac{\partial }{\partial \theta } \mathcal{P}[\psi \overline{\psi }] 
-cy\psi\left(\mathcal{P}[\psi \overline{\psi }]
   - (\psi \overline{\psi })_0\right)
\label{bseqncubic}
\end{equation}
where the rescaled time $\widehat{t}$ is defined in terms of the evolution of
cluster radius in Eq.~(\ref{mean_velocity}).
The tri-linear form of the RHS enables us to compute numerically the motion
within a purely Fourier representation.

\begin{figure}[tbp]
\resizebox{0.9\columnwidth}{!}{\includegraphics*{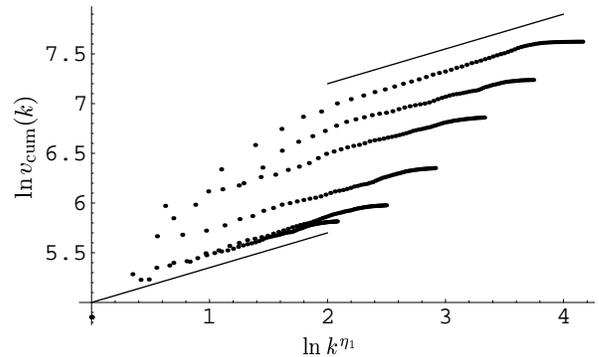}}
\caption{Cumulative contribution to the mean growth velocity plotted against
wavevector as $k^{\protect\eta _{1}}$with logarithmic scales. The data are
(bottom to top) for $\protect\eta _{1}=0.5,0.6,0.7,0.8,0.9,1.0$ and all
exhibit a common power law slope $1/\protect\alpha -1\approx 0.35\pm 0.04$
per the guidelines shown. The result were obtained by numerical integration of
Eq.~(\protect\ref{bseqncubic}) for $c=0$ (periodic strip geometry).}
\label{fig:alphadata}
\end{figure}

In Sec.~\ref{sec:renormalisation} we have seen that the cutoff
dependence of the tip velocity is $v\thicksim K^{\eta _{1}(1/\alpha_{
\text{tip}}-1)}$. This can be compared with the growth rate of the effective
radius, $\frac{dR}{dt}\sim \langle e^{y(\lambda+\bar\lambda)}\rangle =
\langle\psi\bar\psi\rangle = \langle\psi\bar\psi\rangle_0$. So
$\alpha_{\text{tip}}$ can be obtained from the $K$-dependence of
$\langle\psi\bar\psi\rangle_0$ (measured at fixed $R$).  We can obtain it even
from simulations with single $K$: the truncated sum $v_{\text{cum}}(k) =
\sum_{j<k}\left| \psi _{j}\right| ^{2}$ is expected to scale with $k$ in the
same way as the full sum $\langle\psi\bar\psi\rangle_0$ does with $K$,
because the Fourier components far below the cutoff should be insensitive to
the value of $K$.
Figure~\ref{fig:alphadata} shows the measured variation of
$v_{\text{cum}}(k) = \sum_{j<k}\left| \psi _{j}\right| ^{2}$ vs $k^{\eta
_{1}}$: this is expected to exhibit a power law with exponent $(1/\alpha -1)$
and remarkably we obtain $\alpha \approx 0.74\pm 0.02$ with no significant
dependence on $\eta _{1}$ in the range studied.

It is a remarkable success for the Gaussian Theory to have predicted the
completely unexpected insensitivity of $\alpha _{\text{tip}}$ to $\eta $ . 
Whether this result can be truly an exact ``superuniversality'' is another
matter, as certainly the Gaussian value for $\alpha _{\text{tip}}$ is only
approximate and, as we discuss below, matters become more complicated for
large $\eta $.

\section{Breakdown of the DBM model at large $\protect\eta$.}

\label{sec:breakdown}

Sanchez and Sander \cite{sanchez93} first noted that at high enough $\eta
_{0}$ the DBM degenerates because all growth is dominated by the most active
site, showing by direct simulations (at $m=0$) that this happened around $%
\eta _{0}^{c}\approx 4$, a value reinforced by later discussions \cite
{hastings01pre,halsey02} and new data \cite{hastings01prl}. These
discussions are particular to the $m=0$ representation and we believe can be
associated with the degeneration of the moment governing the rate of gain of
cluster mass: this scales with exponent $\tau (\eta _{0})$ which degenerates
to value $\eta _{0}\alpha _{\text{min}}$ when the moment becomes dominated
by the (left) end point $f(\alpha )=0$ of the multifractal spectrum. The
fractal dimension given by $d_{f}=1+\eta _{0}\alpha _{\text{tip}}-\tau (\eta
_{0})$ then degenerates to $d_{f}=1+\eta _{0}\left( \alpha _{\text{tip}%
}-\alpha _{\text{min}}\right) $. If the least screened sites are the tips, $%
\alpha _{\text{tip}}=\alpha _{\text{min}}$, then this also leads to $d_{f}=1$
when $\eta =\eta _{0}^{c}.$

The electrostatic scaling law leads to \textit{earlier} transition in the
behaviour, that is at lower $\eta $, which is also more generic in that it
does \textit{not} depend on growth at some particular value of $m$. This
transition should also limit the applicability of calculating \cite
{hastings01pre,halsey02} exponents perturbatively about $\eta _{0}^{c}$. The
screening transition arises because the moment governing the mean screening
of sites has exponent $\tau (2+\eta _{0})$ which duly appears in the DBM
version Eq.~(\ref{electrostatic-scaling-law-eqn}) of the electrostatic
scaling law, and this moment must hit the end of the $f(\alpha )$ spectrum
before that corresponding to $\tau (\eta _{0})$ discussed above. Once we
have hit this regime, at $\eta _{0}\geq \eta _{0}^{s}$, we have $\tau
(2+\eta _{0})=(2+\eta _{0})\alpha _{\text{min}}$ and the electrostatic
scaling law degenerates to a form which can be rearranged to give 
\begin{equation}
2-\frac{1}{\alpha _{\text{min}}}=\eta _{1}\left( \frac{1}{\alpha _{\text{min}%
}}-\frac{1}{\alpha _{\text{tip}}}\right) .
\end{equation}
This then leads us to choose between two scenarios: either (i) $\alpha _{%
\text{tip}}>\alpha _{\text{min}}$ in which case the behaviour remains
non-trivial, or else (ii) $\alpha _{\text{min}}=1/2$ which means the arms of
the growth are essentially straight and we might suspect self-affine
structure.

\begin{figure}[tbp]
\resizebox{0.9\columnwidth}{!}{\includegraphics*{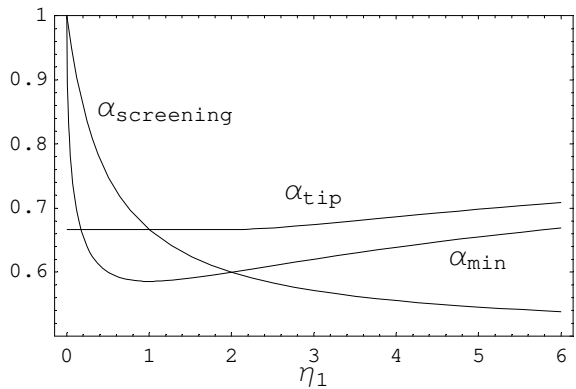}}
\caption{ Singularity exponents reflecting the strength of screening as
calculated from the Gaussian Closure Approximation with $\protect\alpha _{%
\text{tip}}$ set by the electrostatic scaling law. Note that for this theory 
$\protect\alpha _{\text{tip}}>\protect\alpha _{\text{min}}$ meaning that
(within the theory!) the leading tips are not the most active sites. The
screening transition arises when $\protect\alpha _{\text{screening}}$
governing the overall screening hits $\protect\alpha _{\text{min}},$ which
it must subsequently follow.}
\label{fig:alphacalcs}
\end{figure}

The Gaussian Closure Approximation, with $\alpha _{\text{tip}}$ set by the
electrostatic scaling law, has $\alpha _{\text{tip}}>\alpha _{\text{min}}$
for almost all $\eta ,$ leading to scenario (ii) above. Figure~\ref
{fig:alphacalcs} shows the predicted variation of $\alpha _{\text{tip}}$, $%
\alpha _{\text{min}}$ and the value $\alpha _{\text{screening}}$
corresponding to the exponent $\tau (2+\eta _{0}).$ The screening transition
where $\alpha _{\text{screening}}$ hits $\alpha _{\text{min}}$ occurs (it
can be checked exactly) at $\eta _{1}^{s}=2$ corresponding to $\eta
_{0}^{s}=3.$ Beyond this point $\alpha _{\text{tip }}$is no longer quite
constant and stays clear of $\alpha _{\text{min}}$, because of the changed
functional form for $\tau (2+\eta _{0}),$ whilst of course $\alpha _{\text{%
screening}}$ follows $\alpha _{\text{min}}.$%

\begin{figure}[tbp]
\resizebox{0.9\columnwidth}{!}{\includegraphics*{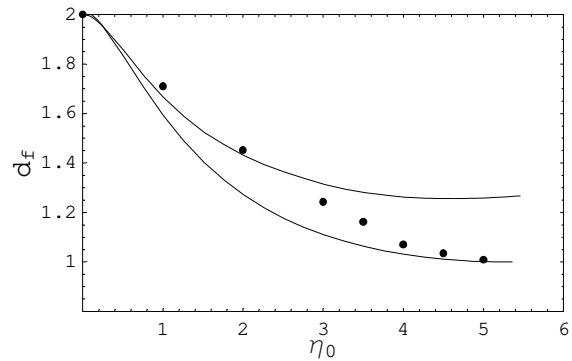}}
\caption{ The fractal dimension for the $m=0$ Dielectric Breakdown Model as
a function of $\protect\eta _{0},$ using the Gaussian Closure Approximation
(curves) compared with published simulation data (points) \protect\cite
{hastings01prl}. The upper curve uses $\protect\alpha _{\text{tip}}$ set by
the Electrostatic Scaling Law, which gives $\protect\alpha _{\text{tip}}>%
\protect\alpha _{\text{min}}$ which is why $d_{f}$ does not approach unity
at the end point $\protect\eta _{0}^{c}.$ The lower curve shows how $d_{f}$
does smoothly approach unity when we force $\protect\alpha _{\text{tip}}=%
\protect\alpha _{\text{min}}.$}
\label{df}
\end{figure}

The corresponding predicted behaviour of the fractal dimension calculated
from $d_{f}=1+\eta _{0}\alpha _{\text{tip}}-\tau (\eta _{0})$ is shown in
Fig.~\ref{df} (upper curve) as a function of $\eta _{0}$ up to $\eta
_{0}^{c}\approx 5.4,$ where it has not fallen to unity because $\alpha _{%
\text{tip}}>\alpha _{\text{min}}$ is maintained. There is some change of
functional form across $\eta _{0}=\eta _{0}^{s}$ but it is scarcely
noticeable graphically. Also shown for comparison (lower curve) is the
behaviour when we force $\alpha _{\text{tip}}=\alpha _{\text{min}}$ instead
of obeying the electrostatic scaling law: in this case the fractal dimension
does smoothly approach unity as $\eta _{0}\rightarrow \eta _{0}^{c},$ but
unfortunately having sacrificed the electrostatic scaling law we cannot see
anything relating to the screening transition. 
The predicted screening transition at $\eta _{0}^{s}=3$ seems to mark a
break in the match to the simulation data of Hastings \cite{hastings01prl}:
below this conforming to the electrostatic scaling law gives the better
agreement, whereas beyond this better agreement comes from forcing $\alpha _{%
\text{tip}}=\alpha _{\text{min}}$. The simplest interpretation is that the
exact answer conforms to both conditions, and it is just their relative
importance which changes around the screening transition.

\section{The Penetration Depth}

\label{sec:penetration-depth}

The multifractal spectrum suggests that the Gaussian approximation is good
in the growth zone, so we have computed as a further test the relative
penetration depth $\Xi $, defined for DLA as the standard deviation $\xi$ of
radius of deposition divided by the effective radius $R$. For DBM more
generally, we have for tractability used as measure the diffusion flux
rather than the local growth rate. 

\begin{figure}[tbp]
\resizebox{0.9\columnwidth}{!}{\includegraphics*{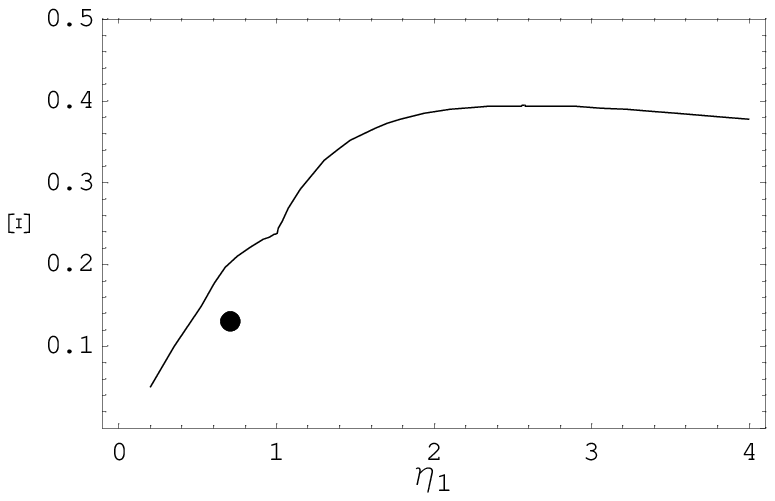}} %
\resizebox{0.9\columnwidth}{!}{\includegraphics*{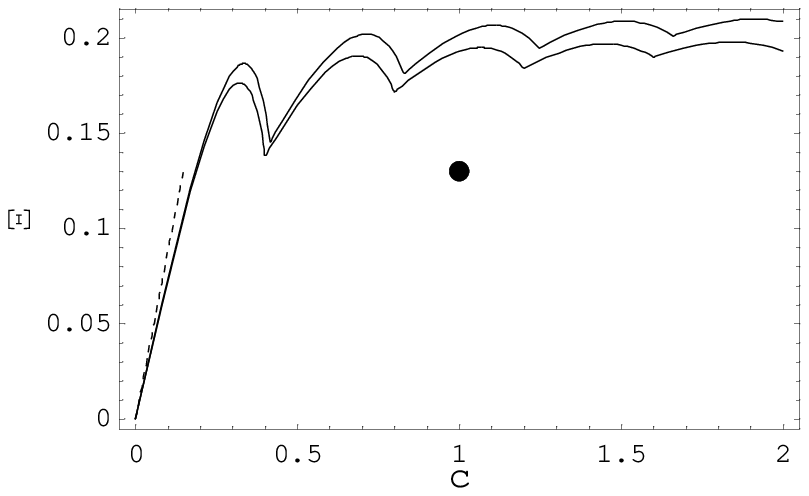}} 
\caption{ Upper panel shows the penetration depth relative to radius, for
radial DBM growth as a function of $\protect\eta _{1}=\protect\alpha \protect%
\eta _{0}$. Lower panel shows the prediction for the DLA case, when growing
in a wedge of angle $2\protect\pi c $, plotted against $c$.  The lower curve
used $\protect\eta _{1}=2/3$ and the upper curve used $\protect\eta
_{1}=0.71$.  The solid point corresponds to measurement for radial DLA, and
the dashed line shows the limiting slope implied by measurements on DLA in a
channel. Note the cusps predicted where $c$ is an integer multiple of
$\frac{\protect\eta _{1}}{1+\protect\eta _{1}}.$ }
\label{fig:xsi}
\end{figure}

The key idea behind the calculation is that we calculate the relative
distortion of the conformal map of the interface
$z(\theta)=Re^{ic\theta}\left(1+\sum_{k>0}w_ke^{ik\theta}\right)$, away from
the circular arc $z_0(\theta)=Re^{i c \theta }$. Following
Ref.~\cite{somfai99}, the squared relative penetration depth is then given by
\begin{equation}
\frac{\xi^2}{R^2} = \int_0^{2\pi}\frac{d\theta}{2\pi}
\left(\Re\frac{z(\theta)-z_0(\theta)}{z_0(\theta)}\right)^2
\end{equation}
which can be expressed in terms of the coefficients $w_k$ as
\begin{equation}
\frac{\xi^2}{R^2} = \frac{1}{2}\sum_{k>0}|w_k|^2 \,.
\end{equation}
From the definition (\ref{logvars}) of $\lambda(\theta)$, the coefficients can
be identified as 
$w_k=\frac{c}{c-k}\left(e^{-\lambda(\theta)}\right)_k$, giving
\begin{equation}
\begin{split}
\frac{\xi ^{2}}{R^{2}} = \; &
\frac{1}{2}
\sum_{k>0}\frac{c ^{2}}{(c -k)^{2}}\int_{0}^{2\pi }\frac{d\theta }{2\pi }
e^{ik\theta }\int_{0}^{2\pi }\frac{d\phi }{2\pi }
\label{xsisquared} \\
&\times e^{-ik\phi }\left( e^{-\sum_{p}\lambda _{p}e^{-ip\theta }-\sum_{p}%
\overline{\lambda }_{p}e^{ip\phi }}\right) \,.
\end{split}
\end{equation}
This expression remains to be averaged over the distribution of cluster
geometries. Appendix~\ref{app:penetration} details the averaging of this under
the Gaussian Closure Approximation, leading to the results shown in Fig.~\ref
{fig:xsi}.

For radial DLA the theory predicts $\Xi =(\xi /R)_{\text{r.m.s.}}=0.20$ in
rather modest agreement with $0.13$ extrapolated from simulations \cite
{somfai99}, and the prediction over a range of DBM parameter is shown in Fig.%
\ref{fig:xsi}. The predicted variation of $\Xi $ for DLA grown in a wedge
is also shown, and in the limit of zero wedge angle the value for the
penetration depth relative to the width of the channel is theoretically $%
0.13 $ compared with $0.14$ measured \cite{somfai000}. What is perhaps more
interesting is the prediction of resonant features which arise when $c$,
the wedge angle relative to $2\pi $, is an integer multiple of $%
\frac{\eta _{1}}{1+\eta _{1}},$ corresponding to integer $k^{\ast },$
because the \emph{idea} of resonant angles in DLA has been much discussed
\cite{kessler98,turkevich85,ball86pa}: this is the first direct and explicit
prediction, which we look forward to seeing tested by simulations.

\section{Growth Fluctuations}

\label{sec:growthfluctuations}

Numerical confidence in the scaling properties of DLA was greatly bolstered by
the idea of an intrinsic but low level of self-organised noise
\cite{ball02pre}, so it is natural to ask if the present theory can address
this.  The simulation studies of noise have rested on tracking the extremal
radius, which is hard to extract from our analytic formulation,  so we have had
to compromise on something more accessible theoretically.  The relative
penetration depth has fluctuations which reflect the differing geometry of the
growth, and for the case of growth in channel we have measured these
fluctuations to be $(\delta \Xi /\Xi)_{\text{simulation}}=0.18$.

The corresponding theoretical calculation is a fairly straightforward
generalisation of the penetration depth calculation itself: we simply
calculate the average square of the expression in Eq.~(\ref{xsisquared}), 
minus the square of Eq.~(\ref{xsiaveraged}) to obtain the variance of $\Xi
^{2}$.  The evaluation of this under the Gaussian Closure Approximation is
detailed in Appendix~\ref{app:penetration}.

The most useful comparison is for the channel, $c \rightarrow 0$, 
for which the penetration depth itself happened to be given rather
accurately by the theory. In this case we obtain the relative variance of
the square of the penetration depth as $\left\langle (\xi
/R)^{4}\right\rangle /\left\langle (\xi /R)^{2}\right\rangle ^{2}-1=1.26$, 
leading to $(\delta \Xi /\Xi)_{\text{theory}}=0.56$ which is substantially
higher than the simulation value. It is clear physically that the
penetration depth comes predominantly from the lowest index modes of
$\lambda$, and this is apparent from our expressions above if we linearise
Eq.~(\ref{xsisquared}). Keeping only $\lambda _{1}$for the channel would
then make $\xi /R$ be the magnitude of a single Gaussian distributed complex
scalar, leading to very similar $\delta \Xi /\Xi $. Thus it seems to be
quite fundamentally the Gaussian form of our closure approximation which
leads to overstatement of the penetration depth fluctuations.

\section{Relation to Cone Angle Theory}

\label{sec:cone-angle-theory}

The angular resonances predicted in the penetration depth turn out to be in
interesting correspondence with part of the earlier Cone Angle Theory (CAT)
of DLA \cite{ball86pa}. In that theory a growing cluster was viewed as having
an identifiable number of major arms $n$, and it was then further supposed
that the growth should be marginally stable with respect to the loss of major
arms through competition for growth. The strongest mode of such competition is
where alternate fingers gain and lose, and the condition for this mode to
marginally stable is in the present notation 
\begin{equation}
\left( \frac{n/c }{2}-1\right) \eta _{0}\alpha =1,
\label{stability-condition-CAT}
\end{equation}
as calculated in ref \cite{ball86pa} for the case $c =1$ and $\eta
_{0}=1$. In the CAT fractional $n$ was presumed an acceptable approximation
and condition (\ref{stability-condition-CAT}) was combined with geometrical
approximations to predict $\alpha $, but here let us focus on the values of
$n$ implied. Using $\eta _{1}=\alpha \eta _{0}$ this gives 
\begin{equation}
\frac{n}{2}=c \frac{1+\eta _{1}}{\eta _{1}},
\end{equation}
so our resonance condition corresponds directly to the case where the number
of marginally stable major arms is an even number - which is of course
required for the alternating mode stability calculation to be strictly
applicable.

\begin{figure}[tbp]
\resizebox{0.9\columnwidth}{!}{\includegraphics*{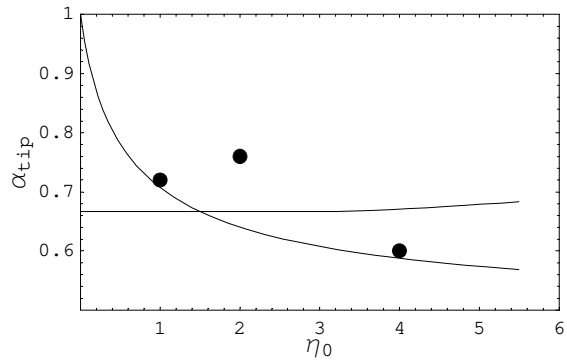}}
\caption{The tip singularity exponent $\protect\alpha _{\text{tip}}$ as a
function of $\protect\eta _{0}$ for the Dielectric Breakdown Model. Rising
curve: Gaussian Closure Approximation; falling curve: Cone Angle Theory;
points: simulation data for $\protect\alpha _{\text{min}}$ from reference 
\protect\cite{sanchez93}.}
\label{fig:alphavalues}
\end{figure}

The CAT was closed in Ref.~\cite{ball86pa} by approximating the cluster as a
solid polygon of $n$ sides, for which $\alpha =\frac{1}{1+2\nu /n}$, 
leading to $\alpha _{\text{CAT}}=\frac{-1+{{\eta }_{0}}+\sqrt{1+6\,{{\eta }
_{0}}+{{{\eta }_{0}}}^{2}}}{4\,{{\eta }_{0}}}$ which is clearly quite
different in principle from the Gaussian Closure prediction of constant
$\alpha _{\text{tip}}$. However it is not easy to distinguish between them on
the basis of previously published DBM data, as shown in
Fig.~\ref{fig:alphavalues}, and unlike GCA the CAT does not predict any other
exponents. 

\section{Discussion}

For DLA and its associated Dielectric Breakdown Models we have shown a
theoretical framework which is complete in the sense that essentially all
measurable quantities can be calculated. This extends to amplitude factors
such as the relative penetration depth for which there is no theoretical
precedent. For the full spectrum of exponents the practical advance over the
Screened Growth Model is the elimination of fitting parameters, and it
remains an open challenge to extend our theory to give quantitatively
credible predictions for the large $\alpha $ part of the spectrum. For the
exponent $\alpha _{\text{tip}}$ we have in the Gaussian approximation a
striking new result that this is predicted constant over a wide range of $%
\eta $, which begs direct confirmation by (expensive) particle-based
simulations. We look forward to addressing this in a following paper.

For the DBM at high $\eta $ we find structure more rich than discussed
hitherto, with a Screening Transition intervening before the upper critical
value $\eta _{0}^{c}$ is reached. Beyond the screening transition the
scenario where $\alpha _{\text{min}}=\alpha _{\text{tip}}$ looks
prospectively solvable (at least in terms of exponents) given the degenerate
form of the Electrostatic Scaling Law which applies. The Gaussian Closure
Approximation leads to the richer but quite possibly misleading scenario $%
\alpha _{\text{min}}<\alpha _{\text{tip}}$, so sorting out the truth of this
inequality would be very interesting.

We have shown that the GCA naturally exhibits angular resonances which are
in interesting correspondence with the ideas of the earlier Cone Angle
Theory. Notably these resonances now have clear predicted consequences such
as we demonstrated for the relative penetration depth, and they can be
explored either by growing in a wedge of variable angle or by varying $\eta $
- so once again behaviour vs $\eta $ is a key probe of our understanding of
the problem. For DLA in particular the best theoretical value of $\alpha _{%
\text{tip}}$ remains $1/\sqrt{2}\approx 0.71$ from the CAT \cite{ball86pa},
but the greater test now lies in the $\eta $ dependence on which CAT and GCA
differ qualitatively.

Within DLA and DBM we look forward to calculating more properties such as
the response to anisotropy, which is fairly readily incorporated into our
equations of motion. The hard part is that in breaking angular symmetry we
can no longer exclude non-zero first cumulants $\left\langle \lambda
_{k}\right\rangle $, and a full matrix of second cumulants, but the
calculation is in principle straightforward. A conceptually more challenging
avenue is to improve on the Gaussian approximation itself which we have used
to obtain explicit theoretical results. Truncating at a cumulant of higher
order than the second is hard, and more seriously it does not correspond to
a positive (semi-)definite probability distribution. An alternative route of
improvement which we are exploring is closure at the level of the full
multifractal spectrum.

Whilst our main use of the equivalences within the class of $(\eta ,m)$
models has been to facilitate calculation, through mapping onto $m=1$, the
particular associated claim that surface tension control is included through 
$m=1$ may prove controversial. This would imply that the scaling properties
of the chaotic viscous fingering regime can be predicted from suitable DBM
simulations. The DBM simulations required are relatively accessible and the
greater difficulty in pursuing this agenda lies in obtaining suitably
calibrated experimental data or accurate direct simulations of fingering out
to high degrees of ramification.

There are possibilities for wider application of ideas in this paper, where
we have formulated DLA and DBM as a turbulent dynamics governed by a complex
scalar field in 1+1 dimensions. Decomposing this field multiplicatively
(through Fourier representation of its logarithm) was the crucial step to
obtain renormalisable equations and theoretical access to the multifractal
behaviour, even though other representations offered equations of motion (%
\ref{bseqncubic}) with weaker non-linearity. It is natural to speculate
whether the same strategy might apply to turbulent problems more widely,
where the key issue appears to be identifying suitable fields to decompose
multiplicatively which are of local physical significance, and subject to
closed equations of motion.

\begin{acknowledgments}
This research has been supported by the EC under Contract No.
HPMF-CT-2000-00800.
\end{acknowledgments}

\appendix

\section{Logarithmic variables}
\label{app:logvar}

In this Appendix we present some details about the logarithmic variables in
Eq.~(\ref{logvars}) and their equation of motion (\ref{logeom}) and
(\ref{mean_velocity}).

We begin by explaining the choice of the analytic form in Eq.~(\ref{logvars}).
In terms of $\omega =\phi +i\theta ,$ in which complex plane the region
exterior to the growth is mapped to a [half] strip, Eq.~(\ref{logvars})
analytically continues to 
\begin{equation}
\frac{dz}{d\omega }=c R\,e^{c \omega }e^{-\sum_{k>0;\text{ }%
k\neq c }\lambda _{k}e^{-k\omega }}.
\end{equation}
Expanding the second exponential factor above to all orders gives 
\begin{multline}
\frac{dz}{d\omega }=c R\,e^{c \omega } \big( 1-\lambda
_{1}e^{-\omega }-(\lambda _{2}-\lambda _{1}^{2}/2)e^{-2\omega }
\\ -(\lambda
_{3}-...)e^{-3\omega }-\dots\big)
\end{multline}
and integrating with respect to $\omega $ gives a conformal map from the
right hand part of the strip to the exterior region of the wedge as
required.
When $c$ is positive integer, periodic boundary conditions put a restriction on the $\lambda$'s: $\lambda_1=0$ when $c=1$, and for the less interesting cases
$c=2,3,\dots$, the prefactor in front of $e^{-c\omega}$ has to vanish.
The Gaussian Closure solutions (\ref{variances}) automatically satisfy
this constraint, because they have $\lambda_k=0$ for $k\le c$.

In the region far from the growth the leading term dominates, giving
$z(\omega)=R\,e^{c \omega }$ which shows the significance of $R$: it is the
apparent radius of the growth (corresponding to $\phi =0$) as seen from far
away.

To obtain the transformed equation of motion (\ref{logeom}), first take the
logarithm of Eq.~(\ref{logvars}), giving $-\lambda (\theta )+\ln R=\ln \left( 
\frac{\partial z}{\partial \theta }\right) +\ln \left( -i/c
e^{-i c \theta }\right) $. Then differentiating both sides with
respect to time (at constant $\theta $) gives 
\begin{equation}
\begin{split}
-\frac{\partial \lambda }{\partial t}+\frac{\partial R}{R\partial t}&=\left( 
\frac{\partial z}{\partial \theta }\right) ^{-1}\frac{\partial }{\partial
\theta }\frac{\partial z}{\partial t}
\\&=\left(c+i\frac{\partial \lambda }{\partial \theta }\right)
\mathcal{P}\left[ (cR)^{-2y}e^{y\left( \lambda +\overline{\lambda }%
\right) }\right] 
\\&~~~~-i\frac{\partial }{\partial \theta }\mathcal{P}\left[
(cR)^{-2y}e^{y\left( \lambda +\overline{\lambda }\right) }\right]
\end{split}
\end{equation}
where we have used $\left| \partial \theta /\partial z\right| =(cR)^{-1}e^{%
\Re\lambda }$ and subsequently the powers of $R$ can be taken outside $%
\mathcal{P}$. It is then trivial to take Fourier components of both sides to
obtain Eqs.~(\ref{logeom}) and (\ref{mean_velocity}), the latter coming
from the zeroth component which we chose to be absent from $\lambda $.

\section{Cumulant expansion}
\label{app:cumulant-expansion}

The Kubo Cumulants for independent variables $X_{i}$ are given in terms of
their moments by $\ln \left\langle e^{\sum_{i}\beta _{i}X_{i}}\right\rangle
=\left\langle e^{\sum_{i}\beta _{i}X_{i}}-1\right\rangle _{c},$ where $\beta
_{i}$ are arbitrary (scalar) parameters \cite{kubo62}.  From this it is well
known that by differentiation with respect to parameters one obtains $%
\left\langle X_{1}e^{W}\right\rangle =\left\langle X_{1}e^{W}\right\rangle
_{c}\left\langle e^{W}\right\rangle $ and $\left\langle
X_{1}X_{2}e^{W}\right\rangle =\left( \left\langle
X_{1}X_{2}e^{W}\right\rangle _{c}+\left\langle X_{1}e^{W}\right\rangle
_{c}\left\langle X_{2}e^{W}\right\rangle _{c}\right) \left\langle
e^{W}\right\rangle $, where $W\equiv \sum_{i}\beta _{i}X_{i}$,  and
similar results for higher moments such as $\left\langle
X_{1}X_{2}X_{3}e^{W}\right\rangle $.

To obtain the renormalised equation of motion (\ref{renormalisedEoM}) from
the unrenormalised equation (\ref{unrenormalisedEoM}) we need first to apply
the above to $\left\langle \overline{\lambda }_{k}e_{k}^{y(\lambda +%
\overline{\lambda })}\right\rangle $. Writing this as $\int \frac{d\theta }{%
2\pi }e^{ik\theta }\left\langle \overline{\lambda }_{k}e^{y(\lambda (\theta
)+\overline{\lambda }(\theta ))}\right\rangle $ allows us to apply the
cumulant identities directly, giving $\int \frac{d\theta }{2\pi }%
e^{ik\theta }\left\langle \overline{\lambda }_{k}e^{y(\lambda (\theta )+%
\overline{\lambda }(\theta ))}\right\rangle _{c}\left\langle e^{y(\lambda
(\theta )+\overline{\lambda }(\theta ))}\right\rangle $ and hence $%
\sum_{j}\left\langle \overline{\lambda }_{k}e_{k-j}^{y(\lambda +\overline{%
\lambda })}\right\rangle _{c}\left\langle e_{j}^{y(\lambda +\overline{%
\lambda })}\right\rangle $.  In the translationally invariant case
considered here only the $j=0$ term survives in the latter summation, 
leading to $\left\langle \overline{\lambda }_{k}e_{k}^{y(\lambda +\overline{%
\lambda })}\right\rangle =\left\langle \overline{\lambda }%
_{k}e_{k}^{y(\lambda +\overline{\lambda })}\right\rangle _{c}\left\langle
e_{0}^{y(\lambda +\overline{\lambda })}\right\rangle $ as required. The
calculation of $\left\langle \lambda _{k-j}\overline{\lambda }%
_{k}e_{j}^{y(\lambda +\overline{\lambda })}\right\rangle =\left(
\left\langle \lambda _{k-j}\overline{\lambda }_{k}e_{j}^{y(\lambda +%
\overline{\lambda })}\right\rangle _{c}+\left\langle \lambda
_{k-j}e_{j-k}^{y(\lambda +\overline{\lambda })}\right\rangle
_{c}\left\langle \overline{\lambda }_{k}e_{k}^{y(\lambda +\overline{\lambda }%
)}\right\rangle _{c}\right) \times \left\langle e_{0}^{y(\lambda +\overline{\lambda 
})}\right\rangle $ proceeds along precisely analogous lines.

\section{Legendre transform of an inverse function}
\label{app:legendre}

In Eq.~(\ref{moments}) we are confronted with a slightly unusual situation:
we wish to find the Legendre Transform $f(\alpha )$ of the function $\tau
(q) $,  that is $f(\alpha )=q\alpha -\tau (q)$ where $\alpha =d\tau /dq$,
given a simple form for the inverse function $q(\tau ).$ Let $g(x)$ be the
Legendre Transform of $q(\tau )$,  that is $g(x)=x\tau -q(\tau )$ where $%
x=dq/d\tau $.  Thus $x=1/\alpha $ and we have $g(1/\alpha )=\tau /\alpha
-q=-f(\alpha )/\alpha $, so the two Legendre Transforms are very simply
related.

The example needed from Eq.~(\ref{moments}) has the form $q(\tau )=1+\tau
+b\tau ^{2}$ leading to $x=1+2b\tau $ and hence $g(x)=-1+b\left( \frac{x-1}{%
2b}\right) ^{2}.$ Then we have $f(\alpha )=-\alpha g(1/\alpha )$ leading
directly to Eq.~(\ref{falpha}).

\section{Calculation of the penetration depth}
\label{app:penetration}

We start from the expression given in Eq.~(\ref{xsisquared}) for the square
of the relative penetration depth without any averaging over clusters.  As
the expression is simple exponential in the $\lambda _{k}$ it is
straightforward to average over a Gaussian distribution leading to 
\begin{widetext}
\begin{equation}
\Xi ^{2}=\left\langle \frac{\xi ^{2}}{R^{2}}\right\rangle =
\frac{1}{2}
\sum_{k\ge0}\frac{c ^{2}}{(c -k)^{2}}\int_{0}^{2\pi }\frac{d\sigma}{%
2\pi }e^{ik\sigma}\left( e^{\sum_{p}S(p)e^{-ip\sigma}}-1\right) .
\label{xsiaveraged}
\end{equation}
Note that the term $-1$ in the integrand makes no difference for $k\not=0$,
but we can include $k=0$ in the outer summation.  This enables us to
rearrange the combination of summation over $k$ and integration,  giving 
\begin{equation}
\Xi ^{2}=\frac{c ^{2}}{2}\int_{0}^{\infty }dx\,xe^{c x}\left(
e^{\sum_{p}S(p)e^{-px}}-1\right) ,  \label{xsiaveragedrearranged}
\end{equation}
as is readily verified upon expanding the exponential $e^{%
\sum_{p}S(p)e^{-px}}$to all orders. The summation in this exponential can be
evaluated in closed form using the variances from Eq.~(\ref{variances}) and
the standard forms $u+u^{3}/3+u^{5}/5+.....=\ln \sqrt{\frac{1+u}{1-u}}$ and $%
u^{2}/2+u^{4}/4+u^{6}/6+....=\ln \sqrt{\frac{1}{1-u^{2}}}$ . This leaves one
numerical quadrature to obtain the results shown in Fig.~\ref{fig:xsi}.

To compute the fluctuations in relative penetration depth we again start
from Eq.~(\ref{xsisquared}) and now average its square to give 
\begin{equation}
\begin{split}
\left\langle (\xi /R)^{4}\right\rangle  =\;\; &%
\frac{1}{4}
\sum_{k>0}\frac{c ^{2}}{(c -k)^{2}}\iint_{0}^{2\pi }\frac{%
d\theta }{2\pi }\frac{d\phi }{2\pi }e^{ik(\theta -\phi )}\sum_{k^{\prime }>0}%
\frac{c ^{2}}{(c -k^{\prime })^{2}}\iint_{0}^{2\pi }\frac{%
d\theta ^{\prime }}{2\pi }\frac{d\phi ^{\prime }}{2\pi }e^{ik^{\prime
}(\theta ^{\prime }-\phi ^{\prime })} \\
& \times \left\langle \exp\left( -\sum_{p}\lambda _{p}e^{-ip\theta }-\sum_{p}%
\overline{\lambda }_{p}e^{ip\phi } -\sum_{p}\lambda
_{p}e^{-ip\theta ^{\prime }}-\sum_{p}\overline{\lambda }_{p}e^{ip\phi
^{\prime }} \right) \right\rangle 
\end{split}
\end{equation}
where averaging the last factor gives \ $\exp\left({\sum_{p}S(p)[e^{ip(%
\phi -\theta )}+e^{ip(\phi -\theta +\theta -\theta ^{\prime })}+e^{ip(\phi
^{\prime }-\theta ^{\prime }+\theta ^{\prime }-\theta )}+e^{ip(\phi ^{\prime
}-\theta ^{\prime })}]}\right)$. One
integration with respect to an absolute angle is now redundant, and the
integrations with respect to $\phi -\theta $ and $\phi ^{\prime }-\theta
^{\prime }$ can be rearranged using the same trick as in calculating $%
\left\langle (\xi /R)^{2}\right\rangle $ previously, leading after some
cancellations to 
\begin{equation}
\begin{split}
\delta \left( \Xi ^{2}\right) ^{2} &=\left\langle \left( \frac{\xi }{R}%
\right) ^{4}\right\rangle -\left\langle \left( \frac{\xi }{R}\right)
^{2}\right\rangle ^{2} \\
&=\frac{c ^{4}}{4}\int_{0}^{\infty }dx\,xe^{c
x}\int_{0}^{\infty }dx'x'e^{c x'}\int_{0}^{2\pi }\frac{d\psi }{2\pi }%
\left( e^{\sum_{p}S(p)[e^{-px}+e^{-p(x-i\psi )}+e^{-p(x'+i\psi
)}+e^{-px'}]}-e^{\sum_{p}S(p)[e^{-px}+e^{-px'}]}\right) .
\end{split}
\end{equation}
Above we used the simplification $\int_{0}^{2\pi }\frac{d\psi }{2\pi }
e^{\sum_{p}S(p)e^{-p(x-i\psi )}}=1$, which follows upon expanding the
exponential where only the leading term survives. This can be further applied
leading to the more compact form $\delta \left(
\Xi ^{2}\right) ^{2}=\int_{0}^{2\pi }\frac{d\psi }{2\pi }\left| \frac{%
c ^{2}}{2}\int_{0}^{\infty }dx\,xe^{c
x}e^{\sum_{p}S(p)e^{-px}}\left( e^{\sum_{p}S(p)e^{-p(x-i\psi )}}-1\right)
\right| ^{2}$. Particularly in order to address the limit of channel
growth, $c \rightarrow 0,$ it is convenient to bypass normalisation
conventions by looking at the relative fluctuations in $\Xi ^{2}$ which are
now given by 
\begin{equation}
\left( \frac{\delta \left( \Xi ^{2}\right) }{\Xi ^{2}}\right) ^{2}=\frac{%
\int_{0}^{2\pi }\frac{d\psi }{2\pi }\left| \int_{0}^{\infty }dx\,xe^{c
x}e^{\sum_{p}S(p)e^{-px}}\left( e^{\sum_{p}S(p)e^{-p(x-i\psi )}}-1\right)
\right| ^{2}}{\left[ \int_{0}^{\infty }dx\,xe^{c x}\left(
e^{\sum_{p}S(p)e^{-px}}-1\right) \right] ^{2}}.
\end{equation}
From this we obtained the result cited in Sec.~\ref{sec:penetration-depth} 
by numerical quadrature.

\end{widetext}

\bibliography{dlaref}

\end{document}